# EUROPEAN HISTORICAL EVIDENCE OF THE SUPERNOVA OF AD 1054 BALKAN MEDIEVAL TOMBSTONES

Miroslav D. Filipović[1*], Miro Ilić[2], Thomas Jarrett[1,3], Jeffrey L. Payne[1], Dejan Urošević[4], Nick F. H. Tothill[1], Patrick J. Kavanagh[5], Giuseppe Longo[6], Evan J. Crawford[1] and Jordan D. Collier[1,3]

[1] *Western Sydney University, School of Science, Locked Bag 1797, Penrith, NSW 2751, Australia*
[2] *Trebinje Astronomical Association, Trebinje, Republic Srpska, Bosnia and Herzegovina*
[3] *University of Cape Town, The Inter-University Institute for Data Intensive Astronomy (IDIA), Department of Astronomy, Private Bag X3, Rondebosch 7701, South Africa*
[4] *University of Belgrade, Faculty of Mathematics, Department of Astronomy, Studentski trg 16, 11000 Belgrade, Republic of Serbia*
[5] *Dublin Institute for Advanced Studies, School of Cosmic Physics, 31 Fitzwilliam Place, Dublin 2, Ireland*
[6] *University Federico II, Department of Physics, Via Cinthia 6, I-80126 Napoli, Italy*

**Abstract**

In a previous work, we establish that the acclaimed 'Arabic' records of SN 1054 from ibn Butlan originate from Europe. Also, we reconstructed the European sky at the time of the event and find that the 'new star' (SN 1054) was in the west while the planet Venus was on the opposite side of the sky (in the east) with the Sun sited directly between these two equally bright objects, as documented in East-Asian records. Here, we investigate the engravings on tombstones (stećci) from several necropolises in present-day Bosnia and Herzegovina (far from the influence of the Church) as a possible European 'record' of SN 1054. Certainly, knowledge and understanding of celestial events (such as supernovae) were somewhat poor in the mid-XI century.

## 1. Introduction

In [1], we noted that supernovae (SNe), such as SN 1054 (a.k.a. M1 or the Crab Nebula), were not well understood in the "high middle ages period". Constantine IX Monomachos, was the Emperor of the Eastern Roman Empire ruling for 4598 days from 1042 A.D. until early 1055 (Julian Calendar). In south-eastern Europe, Science was accorded to doctrine supporting a flat Earth located in the centre of the Universe - the official Christian Church dogma of the time. As the centre of the Solar System, the planets and Sun moved in epicycles around the Earth and nothing ever changed beyond the Moon. To question these well-established laws and defy the Church could meet with serious consequences including arrest and even execution.

---
[*]E-mail: m.filipovic@westernsydney.edu.au

As discussed in [1], appearance of SN 1054 on 4$^{th}$ July 1054 with apparent visibility during the daytime sky for 23 days implies a stunning and colossal historical event. A clearly visible SN event is very rare as many generations will not experience such an event. It is quite extraordinary and impossible to ignore. Simple questions about non-existing direct records in European history are very puzzling and have provoked various searches over the past decades.

As per [2], it is essential to understanding the historical context in which this astronomical event happened. Therefore, we investigated the political, cultural and scientific environment of the time around AD 1054 [1]. As we noted, this was a turbulent period between clergy of the Christian Church in Constantinople and Rome. While differences between the Patriarch in Constantinople and the Archbishop of Rome (the Pope) were not uncommon, this particular period culminated in the Great Schism (the break in communion between what is now the Roman Catholic Church and the Eastern Orthodox Church) which still exists today some ten centuries later [3]!

In [1], we also explored the history of SNe above Europe, noting the evolution of European scientists from astrologers to astronomers that progressed over 518 years, to the time when Tycho's SN 1572 was observed and recorded. However, just 48 years before SN 1054, SN 1006 was visible from Europe, recorded in several European (Italy and Switzerland) chronicles [4-7] and in Arabic texts as noted in [8-10], yet, interestingly, not in any extant East Roman Empire records. It was then several centuries after SN 1054, that two more SNe appeared and were recorded by European astronomers; in 1572 (Tycho's SN) and in 1604 (Kepler's SN) [2].

We used Stellarium software to reconstruct the sky above Constantinople on 16$^{th}$ July 1054 and found that Venus was in the east (in Leo) and SN 1054 was in the west with the Sun (in Gemini) directly between these two equally bright objects. SN 1054 would have been seen above the eastern horizon in the early hours of the morning followed several hours later by sunrise and then Venus, the morning star [1]. That is to say, they appear together on opposite sides of the sky in July of 1054.

In the present paper, we wish to further explore the possibility that some peoples within Europe may also have witnessed the event. Interesting symbols found on tombstones from Bosnia and Herzegovina (BiH) may provide a clue. We discuss these symbols in detail in the following Section 2.

## 2. Tombstones from Radimlja, Hodovo and Zabrđe in Bosnia and Herzegovina

While Christianity may have dominated a significant part of the XI[th] century Europe, there were many remote areas where its influence was either weak (Central and Eastern Europe) or even non-existent (Pomerania (Pomorani) i.e. Poland & Western Russia) [11]. We expect that ordinary people at the time of the 1054 supernova would have been amazed by the appearance of a bright new star in the clear summer sky and would have been keen to somehow record this event. While it is possible the 'new star' was mistaken for a planet (such as Jupiter or Venus), this is unlikely since people in these areas should have been quite familiar with the night sky. One such place, where Christianity was still in its infancy around AD 1054, was the Western and Central Balkan region, which was mainly populated by Serbs. While some of them had recently converted to Christianity, they also retained some of their traditional pagan rituals as part of their everyday life.

As pagans, the Serbs celebrated many gods. They had opposed Christianity for several centuries, most likely because they wanted to preserve their culture and their language. But the mid XI[th] century was a transition period when they obtained some concessions from Constantinople and from Rome, if they would officially convert to Christianity. While most Serbian families celebrated their own Patron Saint's Day (called Slava; these celebrations have been dated back to AD 1018 [12]), they also celebrated wars, nature, summer, spring, food, etc. This was somewhat atypical of many other Christian churches that also were founded at this time. In fact, the decentralised Christian churches from this region did not have a huge significance, and some authors link them to the 'startup' Church called the Bosnian Church, or the Bogomils [13, 14]. While essentially Christian, the Church rituals and practices differed somewhat from those found in Constantinople and in Rome. Therefore, the reaction of the local population to the SN may have been very different to that which occurred in the heart of the Eastern and Western Christian Empires.

Radimlja is a tombstone (stećak, meaning tall, standing thing) necropolis located near Stolac in Bosnia and Herzegovina (BiH; Latitude = 43º05'31.97' N and Longitude = 17º55'26.59' E). These tombstones are monumental medieval graves that lie scattered across the Central and Western Balkans. A number of historians [15-18] date them to around the mid XII[th] century (with the oldest one dating to around 1150 AD) based on a few written epitaphs to local rulers but we note that only 3% of the 60 000 tombstones have useful written epitaphs. However, some of these tombstones most likely originate from earlier periods, as pointed by [18, 19]. The reason behind this suggestion is that most of the tombstones (and especially ones without clear epitaphs) have no typical Christian signs (e.g. the Christian cross is found on only 11 out of 133 tombstones in Radimlja), which points to their pre-Christian origins. Also, [18, 20] has suggested that most of these necropolises were merged and built into previous Antics cemeteries. At the same time [18, p. 89] suggests that the tombstones are positioned in calculated position to follow daily and seasonal path of the Sun.

Nor should one forget that over the past 1000 years many of these tombstones were exposed to 'vandals' who may have wanted to change or destroy these monuments and grave rob. Weathering also played a role, even though all of the tombstones were carved from very hard stone. If this area was populated by a mix of 'half'-Christians, Christians and pagans from the $X^{th}$ to the $XV^{th}$ century they would certainly have been in danger from neighbouring Christian believers.

We investigated the possibility of 14C (radiocarbon) and/or obsidian hydration (OHD) dating these tombstones. Unfortunately, the radiocarbon dating may only be used on organic materials and there is no carbon in stones. Also, during the 'conservation' process of stones in 2017 they have been washed out and OHD method would not return reliable results. Therefore, our tombstones would present enormous challenge to any accurate dating.

The necropolis in Radimlja is particularly interesting in that five tombstones are dedicated to a hunter and his weapons (Figure 1). Other sites, such as the one at Kneževo next to Nevesinje (BiH) and Ljubljenica (Stolac, BiH), also have the same or a similar figure.

Furthermore, we find a resemblance between the hunter on the Radimlja tombstones and the constellation Orion (Figure 2). The most striking feature is the position of a circle between the hunter's right hand and head. We suggest that this circle might pin-point the SN 1054 event. The circle is within ~5º of SN 1054 (based on the Orion constellation as a frame of reference), which we suggest was well within the precision accuracy of a stone mason at that time [21] suggested that it would not be surprising if SN 1054 was interpreted as showing a disk. Alternatively, some historians suggest that circles on these types of tombstones may represent the Sun at the summer solstice, but this is very unlikely as the Sun's position at summer solstice during the X to the XV centuries (when this type of tombstone was made) was ~35º east of the SN 1054 position. So, any carving including the Sun would show it to the left of the raised hand.

Perhaps, this SN 1054 event can be interpreted as a "new Sun"? [18, p. 84] suggests that this particular figure of a hunter symbolises Saint Vitus (also called Vidovdan or Sveti Vid), who was one of the most celebrated Saints in Serbian Christian culture (and was celebrated on Vid's day). The whole Radimlja region is full of toponyms pointing to this saint (exp. Vidovo Polje, Vidostica, Vidoška Utvrda, Vidoštak, etc) [18, p. 85]. It is interesting that the Serbian Orthodox Church celebrates Vidovdan on $15^{th}$ June ($28^{th}$ June in the Gregorian calendar), quite close to the start of the SN 1054 event. However, there is no evidence that the local population celebrated this Saint in the mid $XI^{th}$ century: the earliest records of Saint Vitus celebrations among the Serbs only date to the $XIX^{th}$ century.

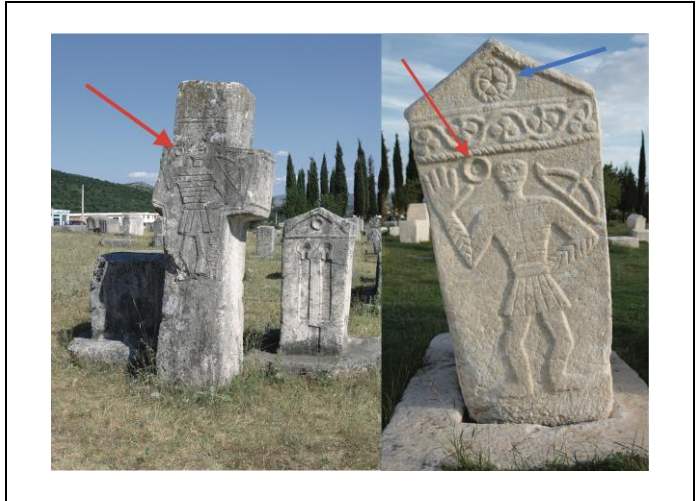

**Figure 1.** Tombstones (stećci) from Radimlja (Bosnia and Herzegovina), depicting a hunter (Saint Vitus and the constellation Orion) with a circle (see the red arrow) between the right hand and head, possibly depicting the SN 1054 event. The blue arrow on the right-hand photograph points to the circular rosetta sign that represents the Sun. The tombstone on the left-hand side is made in a cross-like shape, indicating the influence of early Christianity (photographs: Miro Ilić).

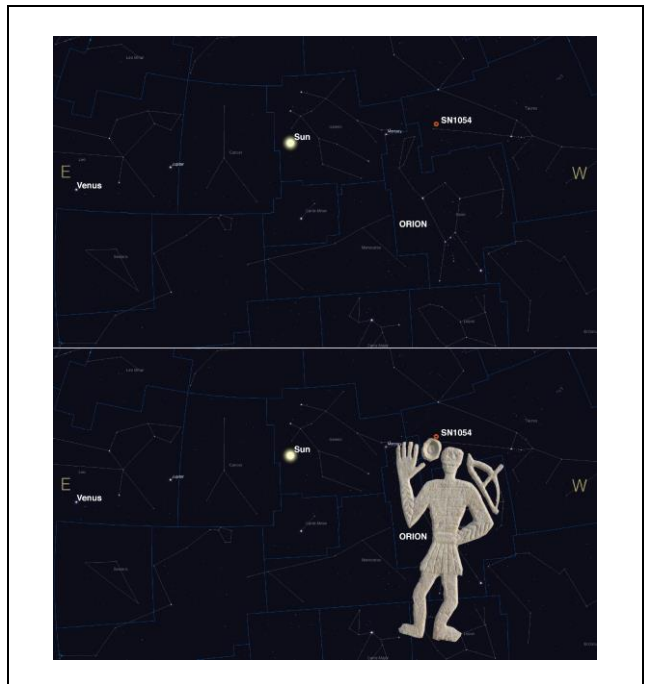

**Figure 2.** The sky above Radimlja (Bosnia and Herzegovina) on 28 June 1054 with the hunter overlaid, as a depiction of the constellation Orion. 4 July 1054 is the most likely date of the explosion of SN 1054 and was close to the summer solstice and day when Christians celebrate Saint Vitus Day.

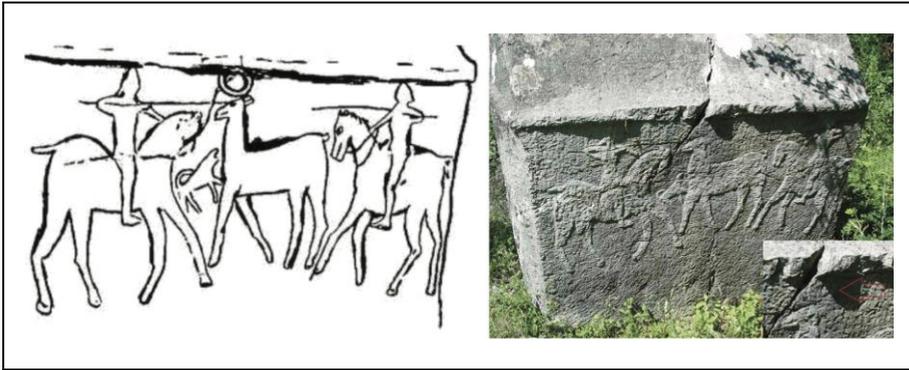

**Figure 3.** Tombstone from Hodovo depicting a deer (the constellation Orion) with a circle above its head (SN 1054), a dog (Canis Major) and two horsemen (representing Taurus and (what is today known as) Monoceros or alternatively, Saint George and Saint Elijah). The drawing on the left is after Wenzel [22] and figure on the right is from http://stecci.org/ (©Pierre Auriol 2005). The indented image bottom right shows remaining parts of the circle, marked in red (also after Wenzel [22]).

Very influential and insightful [18] has openly considered a connection between astral symbols on tombstones and the positions of celestial bodies at the time of a person's death, while [23] showed several such examples in nearby necropolises (e.g. at Ponor, Lokvičići and Cista).

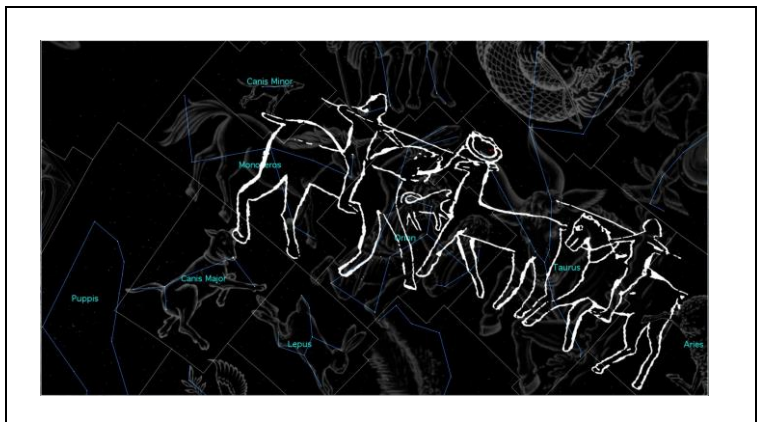

**Figure 4.** The tombstone from Hodovo plotted on top of a sky map by Hevelius.

Viduša suggests that the Sun is actually represented on these particular tombstones by a different circle (rosetta) that is positioned on the top and divided into the six parts, representing the three months of spring and three months of summer [23, 24]. Also, he argues that the raised right hand of the hunter is a sign of the summer solstice, compared to various other hand positions that point to winter, autumn or spring. Among a number of different explanations, Lovrenović argues that the oversized hand represents an act of being surprised (with SN 1054?) [18]. Could the position of this circle, together with Orion being Saint Vitus, be a ciphered (if not explicit) message about the SN 1054 event? We also found it interesting that these tombstones with the hunter and star between the head and hand are all orientated towards the west - the same direction as SN 1054.

Also dating from the mediaeval period and some 11.6 km north of Radimlja there is another necropolis named Hodovo where we also found an interesting tombstone (Figure 3). This particular tombstone was first described by [25] and later in [22] and [24, 26]. While some deterioration is obvious, we can still see five major figures: a deer, a dog, two horsemen and a circle above the deer's head. This is the only tombstone with a composition that includes the same circle as seen in Radimlja. All other tombstones with similar motives are notably without a circle. While the precise dating of this tombstone is difficult, we can safely place it before the eleventh or twelfth century because it does not contain any typical Christian motifs (such as crosses, for example). We suggest that the circle on this tombstone could also represent SN 1054, with the deer being 'Orion' and the dog Canis Major. While these three figures can nicely fit the sky and star positions, the two horsemen might represent (what is today known as) Monoceros and Taurus - with their horns being 'spears' (Figure 4). But perhaps the two horsemen who are pointing their spears towards the deer might also be two famous saints who are usually portrayed riding horses, Saint George (Đurđevdan, celebrated on 23$^{rd}$ April) and Saint Elijah (Ilija, celebrated on 20$^{th}$ July). The mid-point between these two saints' day celebrations is 6$^{th}$ June, which is very close to the date of the SN 1054 event [25, p. 76]. Again, as in the case of Saint Vitus, we do not have firm evidence or documents that can confirm the above hypothesis. One shouldn't forget that the deer is regarded as a holy animal in almost all pre-Christian population from this region [27]. Some authors are suggesting that a deer might be an astral symbol that takes dead to divine sky and renewal of life. But, in early Christian beliefs, deer is seen as a symbol of soul.

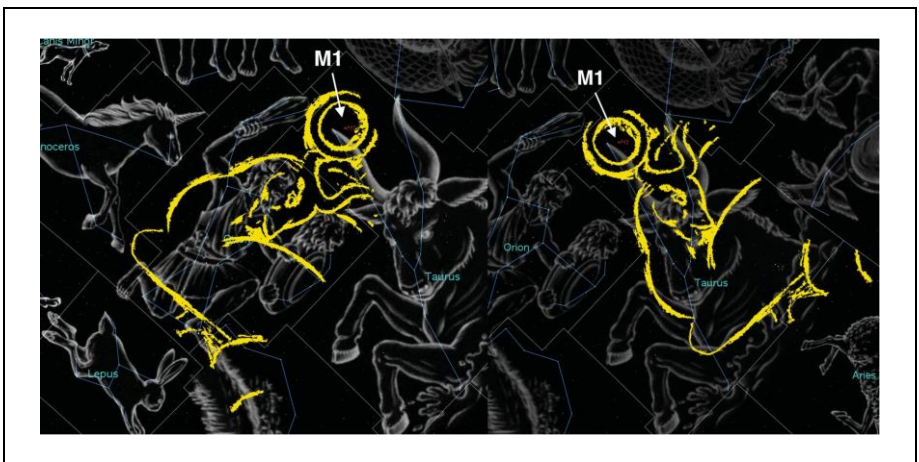

**Figure 5.** The tombstone from Zabrđe plotted on top of a sky map by Hevelius. We suggest that it maybe depicting a deer (the constellation Taurus or Orion) with a circle on top of its horn representing the SN 1054 (M1) event [22, p. 277].

Just few kilometres east of Radimlja, there is another necropolis called Zabrđe (Stolac, BiH; Latitude = 43º06'37.47" N and Longitude = 18º01'09.98" E), where there is another striking tombstone, shown in Figure 5 (and also discovered by Wenzel [22, p. 277, Fig. 20]. This has a deer, with the same circle above one of the horns, and nicely matching the position of SN 1054 (Figure 5).

We also examined photographs of >60 000 tombstones from the central Balkan region. In our study, we specifically focused on a rare seven tombstones with unique but simple ring(s) that were engraved on the hard rocks and found at three different graveyards in central Bosnia and Herzegovina. In all cases, apparently intentional positioning of these rings with respect to the rest of the tombstone features was evident and obvious. All of the engravings on these tens of thousands of tombstones were purposely, artistically and meaningfully made. There were tens-to-hundreds of the same or similar tombstones depicting a hunter, but without the essentially important circle as seen in the Radmilja ones. Purposely and carefully made circles on these tombstones made them very special compared to the rest of the sample.

Furthermore, we noticed numerous other similar ring-like features on other tombstones that also showed artistic representations of the Sun and Moon, and even other celestial bodies and daily life happenings. However, we found that it was the positioning of these specific rings from all three necropolises that could be interpreted as connected to SN 1054.

Probably the best overview of the astral signs variety on these medieval tombstones could be found in [28]. There, very clear indication about this ring-like feature on Radimlja tombstones is connected with a possible 'new Sun' and symbols of possible lightning. According to the same study, early-Christian (old Slavic or post-pagan) population of this region had a very dedicated spiritual connection (admiration, idolatry) with the Sun and to some extent - the Moon. One would certainly expect that a 'new Sun with lightning' will also have significant importance in their daily pagan lives and those rituals would be translated on to the burial sites.

Finally, the Radimlja hunter is similar to various Paleolithic petroglyphs and a wide variety of rock art from non-Western cultures around the world, but with one major difference: the circle at that specific position is unique to these seven tombstones. Therefore, we suggest that these specific tombstone rings are intentional and are not random features nor are they associated with the conventions of the Sun, Moon and planets.

## 3. Concluding remarks

We further examine the records of European history and culture from around 1054 to ask whether what we see is consistent with SN 1054 having been seen in the skies above Europe and having had an impact on the peoples who saw it. Certainly, there are no precise and indisputable European records of SN 1054 that are comparable to the ones from the East-Asian countries. While there is no doubt that most (if not all) of the historical records around SN 1054 event suffer from various biases (from which temporal coincidence is the most dominant), we still present here plausible explanation of some European records on tombstones from Bosnia and Herzegovina that could relate directly to the SN 1054 event. It would be astonishing to see such an event in the daytime sky, and equally astonishing to simply ignore it. It is quite plausible that it was in fact recorded but hidden in plain sight. In our next paper, we will explore another possible way SN 1054 was recorded in Constantinople by means of a cipher.

**Acknowledgment**

We thank Olivera Bjekić, Milan S. Dimitrijević, Branko Simonović, Velibor Velović, Petko Nikolić Viduša and the Western Sydney University Library team led by Linda Thornley for valuable help in reading and discussing various ideas regarding this study. We also thank Aleksandar Zorkić, Ain De Horta and Darren Maybour for technical help in preparing the figures shown here. We acknowledge that this paper made use of the Stellarium software (http://www.stellarium.org/).

**References**

[1] M.D. Filipović, J.L. Payne, J. Thomas, N.F.H. Tothill, D. Urošević, P.J. Kavanagh, G. Longo, E.J. Crawford, J.D. Collier and M. Ilić, Eur. J. Sci. Theol., **17(3)** (2021) 147-160.
[2] F.R. Stephenson and D.A. Green, *Historical supernovae and their remnants*, in *Historical supernovae and their remnants*, International series in astronomy and astrophysics, 5, Clarendon Press, Oxford, 2002, 89.
[3] F. Cross and E. Livingstone, *The Oxford Dictionary of the Christian Church*, Oxford University Press, Oxford, 2005, 1800.
[4] F.R. Stephenson and D.A. Green, Astr. Soc. P., **342(12)** (2005) 63-70.
[5] P.F. Winkler, High. Astron., **14(8)** (2007) 301-302.
[6] P.F. Winkler, Prog. Theor. Phys. Supp., **169** (2007) 150-156.
[7] F.R. Stephenson, Astronomy and Geophysics, **51(5**) (2010) 5.27-5.32.
[8] B.R. Goldstein, Astron. J., **70(1)** (1965) 105.
[9] R. Neuhäuser, C. Ehrig-Eggert and P. Kunitzsch, Astron. Nachr., **338(1)** (2017) 19-25.
[10] R. Neuhäuser, D.L. Neuhäuser, W. Rada, J. Chapman, D. Luge and P. Kunitzsch, Astron. Nachr., **338(1)** (2017) 8-18.
[11] W. Buchholz, *Pommern*, Siedler, München, 1999, 366.
[12] P. Vlahovic, Slavenski etnograf, **20** (1968) 120-128.
[13] J. Fine and J. Van Antwerp Fine, *The Late Medieval Balkans: A Critical Survey from the Late Twelfth Century to the Ottoman Conquest*, ACLS Humanities E-Book, University of Michigan Press, United States, 1994, 41.
[14] T. Stoianovich. *Balkan Worlds: The First and Last Europe*, Taylor & Francis, New York, 2015, 142.
[15] M. Vego. *Stećci. Božanska igra brojki i slova*, Zbornika srednjovjekovnih natpisa Bosne i Hercegovine, Sarajevo, 1962, 1-4.
[16] B. Purgarić-Kužić, Radovi, **28(1)** (1995) 242.
[17] R. Trako, Socijalna ekologija, **20(1)** (2011) 71.

[18] D. Lovrenović, *Stećci: bosansko i humsko mramorje srednjeg vijeka*, Biblioteka Cicero, Naklada Ljevak, 2013, 225-232.
[19] Š. Bešlagić, *Stećci - kultura i umjetnost*, Biblioteka Kulturno nasljeđe, IRO 'Veselin Masleša', OO Izdavačka djelatnost, Sarajevo, 1982, 60.
[20] N. Miletić, *Umetnost na tlu Jugoslavije, Stećci*, Izdavacki zavod Jugoslavija, Beograd, 1982, 22.
[21] G.W. Collins II, W.P. Claspy and J.C. Martin, Publ. Astron. Soc. Pac., **111(7)** (1999) 871-880.
[22] M. Wenzel, *Ukrasni Motivi Na Stećcima: Bibl. Kulturno nasljeđe*, Serbian translation, N. Bibl. Kulturno nasljede, Veselin Masleša, Sarajevo, 1965, 391.
[23] I. Alduk, Contributions to the History of Art in Dalmatia, **42(1)** (2011) 161-186.
[24] P. Viduša, *Petroglifi i stećci*, Mystic Book, Kitchener Canada, 2015, 7-47.
[25] M. Wenzel, J. Warburg Courtauld, **24(1/2)** (1961) 89-107.
[26] P. Viduša, *Geneza stećaka i dvovjerje krstjana*, Pesić i sinovi, Beograd, 2018, 43.
[27] I. Mužić, Starohrvatska prosvjeta, **3(36)** (2009) 315-349.
[28] D. Vidović, *Simbolične predstave na stećima*, II, Naše starine, Sarajevo, 1954, 119-127.